\documentclass[12pt,twoside]{article}

\def\sun{\odot}

\def\sun{\hbox{$\odot$}}
\date{}
\usepackage {graphicx} 
\def\apj{ApJ\,  }

\def\mnras{MNRAS\,  }

\begin{document}
\centerline{\bf Adv. Studies Theor. Phys, Vol. x, 200x, no. xx,
xxx - xxx}

\centerline{}

\centerline{}
\centerline 
{\Large
{\bf 
A right and  left truncated
gamma 
}
}
\centerline 
{\Large
{\bf 
distribution with application  
to  the stars
}
}
\centerline{}

\centerline{\bf {L. Zaninetti}}

\centerline{}

\centerline{Dipartimento  di Fisica ,}

\centerline{Universit\`a degli Studi di Torino,}

\centerline{via P. Giuria 1,  10125 Torino, Italy}

\begin{abstract}
The gamma density function is usually  defined  
in interval between zero and infinity. 
This paper introduces an upper and a lower 
boundary to this distribution.
The parameters which  characterize  the truncated gamma
distribution are evaluated.
A statistical test is  performed on two samples of stars.
A comparison
with the lognormal and the four power law distribution is made.
\end{abstract}

\section{Introduction}

A probability distribution function  (PDF)
which  models a given  physical  variable
is  usually defined  in the interval $0 \leq x < \infty$.
As an example  the exponential , the gamma, the lognormal,
the  Pareto and  the Weibull PDFs are defined in such
interval, see \cite{evans}.
We now briefly review the status of the research
on the  truncated  gamma distribution (TG).
A first attempt to deduce the  parameters  of a TG
can be found  in \cite{Chapman1956},
\cite{BaikunthNath1975}
derived  the minimum variance unbiased 
estimate of the reliability function
associated with the TG distribution  
which is right truncated,
\cite{Hegde1989,univariate1}
estimated   the parameters of a TG
distribution over $0 \leq x <t$,
adopting the maximum likelihood estimator(MLE),
\cite{Philippe1997}
studied the properties of TG distributions 
and  derived the 
simulation algorithms which dominate 
the standard algorithms for these
distributions,
\cite{Coffey2000} considered a doubly-truncated gamma
random variable restricted by both a
lower (l) and upper (u) truncation.

On adopting an astronomical  point of view
the left truncation is connected
with the minimum mass of a star, 
$\approx 0.02 M_{\sun} $
and the right  truncation  with
the maximum mass of a star, $\approx 60  M_{\sun} $,
see  \cite{Kroupa2013}.
This paper first review the gamma PDF, 
introduces the right and  left truncated  gamma PDF
and  finally analyzes two  samples
of  stars  and  brown dwarfs (BD).

\section{The various gamma distributions}

This  Section  reviews the gamma PDF, introduces
the truncated gamma PDF and   analyzes  the data
of  two  astronomical samples.

\subsection{The gamma distribution}

Let $X$ be a random variable taking values $x$ in the interval
$[0, \infty ]$; the {\em gamma } PDF  is
\begin {equation}
f(x;b,c) =
\frac {
 \left( {\frac {x}{b}} \right) ^{c-1}{{\rm e}^{-{\frac {x}{b}}}}
}
{
b\Gamma  \left( c \right)
}
\label{gammastandard}
\end {equation}
where
\begin{equation}
\mathop{\Gamma\/}\nolimits\!\left(z\right)
=\int_{0}^{\infty}e^{{-t}}t^{{z-1}}dt
\quad ,
\end{equation}
is the gamma function,
    $b>0$   is the scale
and $c>0$   is the shape,
see  formula (17.23) in \cite{univariate1}.
Its expected value   is
\begin{equation}
E(x;b,c)=  bc  \quad,
\end{equation}
and its variance,
\begin{equation}
Var (x; b,c) =
b^2 c
 \quad.
\end{equation}
The mode  is at
\begin{equation}
m(x;b,c)=bc-b \quad when \, c>1 \,  \quad.
\end{equation}
The distribution function   (DF)  is
\begin{equation}
DF(x;b,c) =
\frac{ \gamma (c,\frac{x}{b})}
     {\Gamma  \left( c \right)}
\quad ,
\end{equation}
where
\begin{equation}
\mathop{\gamma\/}\nolimits\!\left(a,z\right)=\int_{0}^{z}t^{{a-1}}e^{{-t}%
}dt,
\end{equation}
is the  lower incomplete gamma function,
see
\cite{Abramowitz1965,NIST2010}.
The two parameters can be estimated by
matching the moments
\begin{equation}
b=\frac
{s^2}
{
\bar{x}
}
\label{bvalue}
 \end{equation}
\begin{equation}
c =
(
\frac
{
{ \bar{x}}
}
{s})^2
 \quad,
 \label{cvalue}
\end{equation}
where $s^2$ and  $\bar{x}$ 
are the sample 
variance and the sample mean.
More details can be found in  \cite{evans}.

\subsection{The truncated  gamma distribution}

Let $X$ be a random variable taking
values $x$ in the interval
$[x_l, x_u ]$;
the truncated gamma  (TG) PDF  is
\begin {equation}
f(x;b,c,x_l,x_u) =
k\; \left( {\frac {x}{b}} \right) ^{c-1}{{\rm e}^{-{\frac {x}{b}}}}
\label{gammatruncated}
\end {equation}
where  the constant  $k$
is
\begin{equation}
k =
\frac{c}
{
b\Gamma  \left( 1+c,{\frac {x_{{l}}}{b}} \right) -b\Gamma  \left( 1+c,
{\frac {x_{{u}}}{b}} \right) +{{\rm e}^{-{\frac {x_{{u}}}{b}}}}{b}^{-c
+1}{x_{{u}}}^{c}-{{\rm e}^{-{\frac {x_{{l}}}{b}}}}{b}^{-c+1}{x_{{l}}}^
{c}
}
\quad  ,
\label{constant}
\end {equation}
where
\begin{equation}
\mathop{\Gamma\/}\nolimits\!\left(a,z\right)=\int_{z}^{\infty}t^{{a-1}}e^{{-t}%
}dt,
\end{equation}
is the  upper incomplete gamma function,
see
\cite{Abramowitz1965,NIST2010}.
Its expected value   is
\begin{equation}
E(b,c,x_l,x_u)
=
-{b}^{2}k \left( -\Gamma  \left( 1+c,{\frac {x_{{l}}}{b}} \right) +
\Gamma  \left( 1+c,{\frac {x_{{u}}}{b}} \right)  \right)
\quad .
\label{meangammatruncated}
\end{equation}
The mode  is at
\begin{equation}
m(x;b,c,x_l,x_u)=bc-b \quad when \, c>1 \,  \quad,
\end{equation}
but in order to exist  the  inequality
$x_l < m < x_u$
should be
satisfied.
The distribution 
function    is
\begin{eqnarray}
DF(x;b,c,x_l,x_u)
=   &    \nonumber  \\
k \left( b\Gamma  \left( 1+c,{\frac {x_{{l}}}{b}} \right) -b\Gamma
 \left( 1+c,{\frac {x}{b}} \right) +{{\rm e}^{-{\frac {x}{b}}}}{b}^{-c
+1}{x}^{c}-{{\rm e}^{-{\frac {x_{{l}}}{b}}}}{b}^{-c+1}{x_{{l}}}^{c}
 \right)
\quad  .
\label{dftruncatedgamma}
\end{eqnarray}
A random number generation can be implemented
by solving for $x$ the following nonlinear equation
\begin{equation}
DF(x;b,c,x_l,x_u) - {\bf R} =0
\quad  ,
\label{randomgammatruncated}
\end{equation}
where we have 
a pudendum number generator giving 
random  numbers ${\bf R}$ between zero and one,  
see  \cite{Kahaner1989}.
A simple derivation of the lower and upper boundaries gives 
\begin{equation}
\tilde{x_l}= minimum ~of~sample \quad \tilde{x_u}= maximum ~of~sample
 \quad.
 \end{equation}
A first approximate derivation of   
$\tilde{b}$ and
$\tilde{c}$ 
is through the standard estimation of parameters 
of the gamma distribution.
We compute the $\chi^2$
with these first values of 
$\tilde{b}$ and
$\tilde{c}$ and we search a numerical couple 
which gives the minimum $\chi^2$.
The $\chi^2$
is computed
according to the formula
\begin{equation}
\chi^2 = \sum_{i=1}^n \frac { (T_i - O_i)^2} {T_i},
\label{chisquare}
\end {equation}
where $n  $   is the number of bins,
      $T_i$   is the theoretical value,
and   $O_i$   is the experimental value represented
by the frequencies.
The  merit function $\chi_{red}^2$
is  evaluated  by
\begin{equation}
\chi_{red}^2 = \chi^2/NF
\quad,
\label{chisquarereduced}
\end{equation}
where $NF=n-k$ is the number of degrees  of freedom,
$n$ is the number of bins,
and $k$ is the number of parameters.
The goodness  of the fit can be expressed by
the probability $Q$, see  equation 15.2.12  in \cite{press},
which involves the degrees of freedom
and the $\chi^2$.
The Akaike information criterion
(AIC), see \cite{Akaike1974},
is defined by
\begin{equation}
AIC  = 2k - 2  ln(L)
\quad,
\end {equation}
where $L$ is
the likelihood  function  and 
$k$  the number of  free parameters
in the model.
We assume  a Gaussian distribution for  the errors
and  the likelihood  function
can be derived  from the $\chi^2$ statistic
$L \propto \exp (- \frac{\chi^2}{2} ) $
where  $\chi^2$ has been computed by
Equation~(\ref{chisquare}),
see~\cite{Liddle2004}, \cite{Godlowski2005}.
Now the AIC becomes
\begin{equation}
AIC  = 2k + \chi^2
\quad.
\label{AIC}
\end {equation}

\subsection{Data analysis}

\begin{figure*}
\begin{center}
\includegraphics[width=10cm]{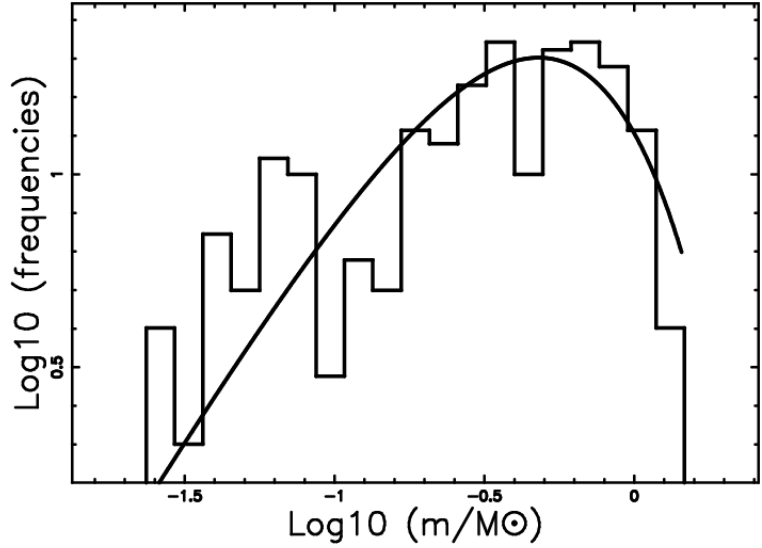}
\end{center}
\caption
{
Logarithmic histogram   of  mass distribution
as  given by   NGC 6611 cluster data (207 stars + BDs)
with a superposition of the TG    distribution
when the number of bins, $n$, is 12,
$c$   = 1.287 ,
$b$   = 0.372 ,
$x_l$ = 0.019
and
$x_u$  =1.36.
Vertical and horizontal axes have logarithmic scales.
}
\label{gamma_tronc_log_ngc6611}
\end{figure*}

\begin{figure*}
\begin{center}
\includegraphics[width=10cm]{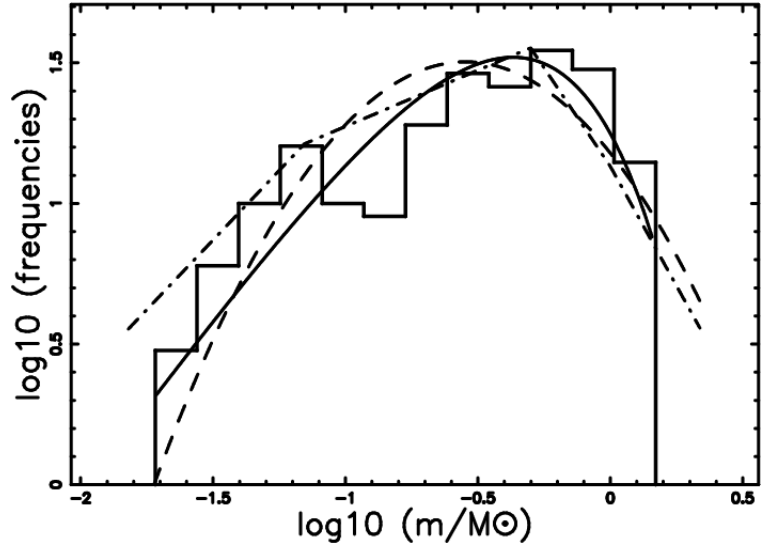}
\end{center}
\caption
{
Histogram (step-diagram)  of  mass distribution
as  given by   NGC 6611 cluster data (207 stars + BDs)
with a superposition of the left
TG    distribution (full line),
the lognormal  (dashed),
and the four  power laws (dot-dash-dot-dash).
Vertical and horizontal axes have logarithmic scales.
}
\label{gamma_tronc_log_3_ngc6611}
\end{figure*}

A first test  is performed  
on   the low-mass initial mass function 
in the young cluster NGC 6611,
see  \cite{Oliveira2009}.
Table
\ref{chi2valuesngc6611}  shows the values of  
$\chi_{red}^2$, 
the
AIC,  the probability $Q$, 
of  the astrophysical fits and the
results of the K-S test, 
the maximum  distance, $D$, between the theoretical
and the astronomical  DF
as well the  significance  level  $P_{KS}$ ,
see \cite{Kolmogoroff1941,Smirnov1948,Massey1951,press}.
Figure  \ref{gamma_tronc_log_ngc6611}  shows the fit
with  the  TG  distribution
of  NGC 6611
and Figure  \ref{gamma_tronc_log_3_ngc6611}
visually compares the three types of fits for NGC 6611.

\begin{table}[ht!]
\caption
{
Numerical values of
NGC 6611 cluster data (207 stars + BDs).
The  number of  linear   bins, $n$, is 20.
}
\label{chi2valuesngc6611}
\begin{center}
\begin{tabular}{|c|c|c|c|c|c|c|}
\hline
PDF       &   parameters  &  AIC  & $\chi_{red}^2$
& $Q$  &  D &   $P_{KS}$  \\
\hline
lognormal &  $\sigma$=1.029,$\mu_{LN} =-1.258 $  &  71.24&
3.73    & $1.3\,10^{-7}$  &  0.09366 &  0.04959 \\
\hline
gamma & b=0.248  ,c = 1.717 & 62.83 & 3.26   & 3.15\,$10^{6}$
& 0.109  & 0.0124   \\
\hline
truncated ~gamma & b=0.372 ,c =1.287
& 52.34  & 2.77    & 0.00017  & 0.09  & 0.061   \\
~  & $x_l$=0.019, $x_u$=1.46
& ~     & ~      &  ~     &~       & ~        \\
\hline
four
    & Eqn.(59)
& 81.39  & 5.18   &  $2.41\,19^{-9}$   & 0.12514     &  $ 2.72\,10^{-3}$  \\
power~ laws  & in~Zaninetti~2013
&  ~            &   ~  & & & \\
\hline
\end{tabular}
\end{center}
\end{table}

\begin{figure*}
\begin{center}
\includegraphics[width=10cm]{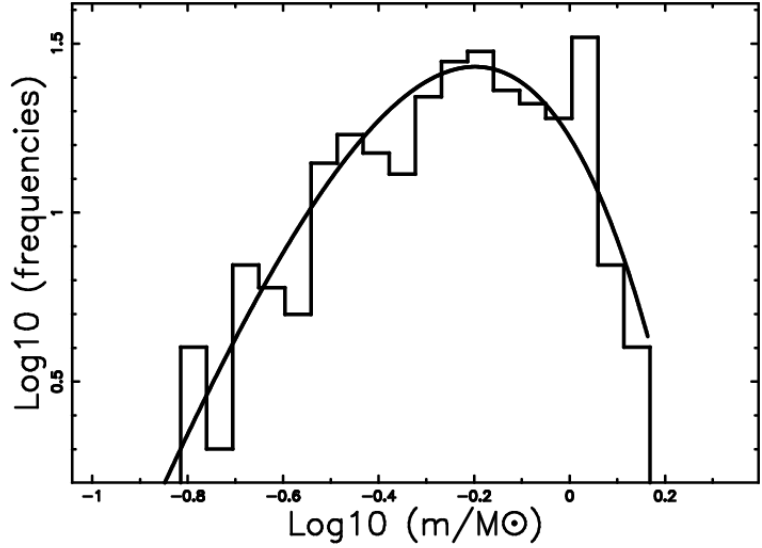}
\end{center}
\caption
{
Logarithmic histogram   of  mass distribution
as  given by   NGC 2362 cluster data (272 stars)
with a superposition of the
TG    distribution
when the number of bins, $n$, is 12,
$b= 0.161$,
$c= 3.933$,
$x_l$= 0.12
and
$x_u$  =1.47~.
Vertical and horizontal axes have logarithmic scales.
}
\label{gamma_tronc_log_ngc2362}
\end{figure*}

\begin{figure*}
\begin{center}
\includegraphics[width=10cm]{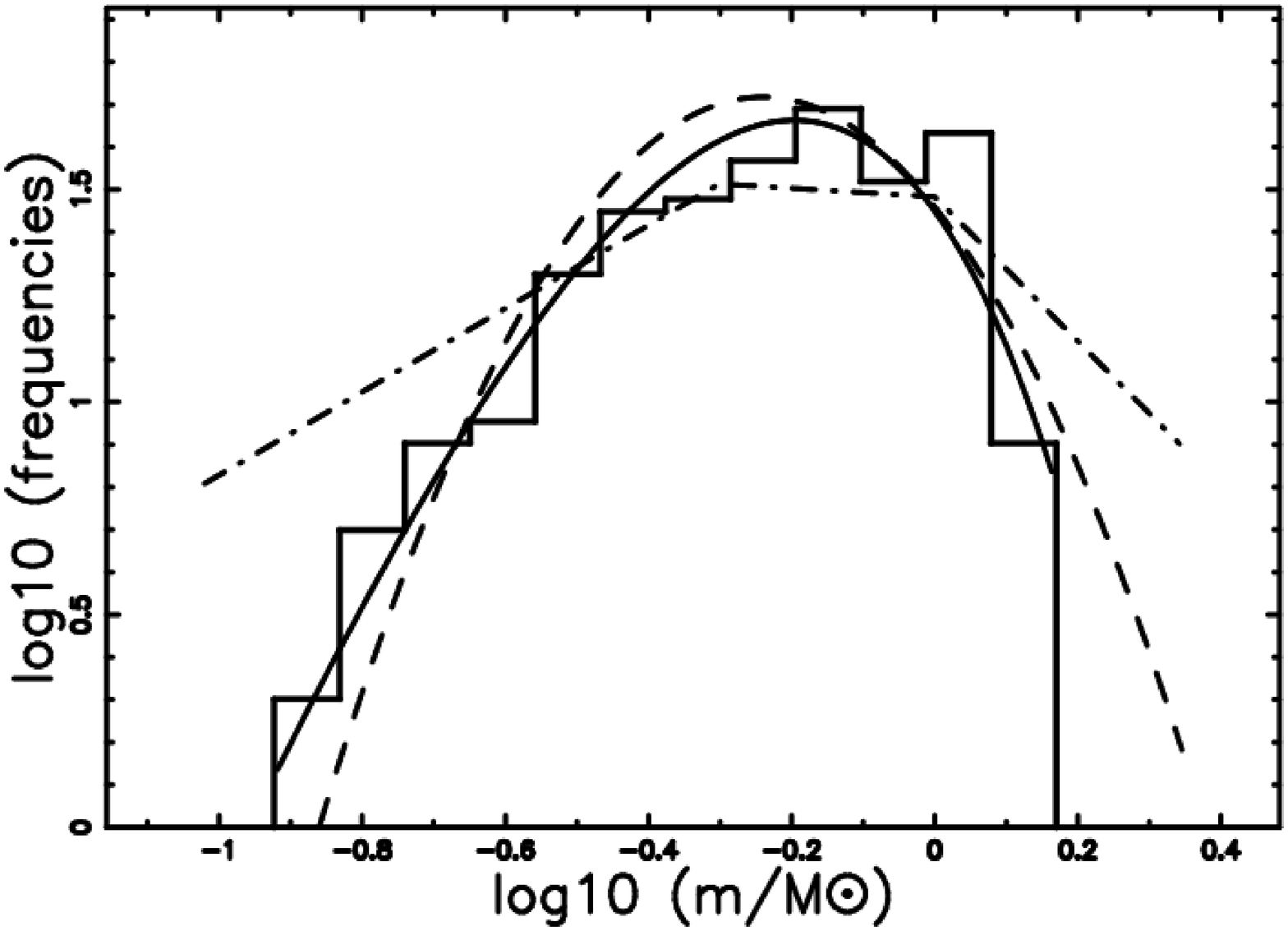}
\end{center}
\caption
{
Histogram (step-diagram)  of  mass distribution
as  given by   NGC 2362 cluster data (272 stars)
with a superposition of the left
TG    distribution (full line),
the lognornal  (dashed),
and the four  power laws (dot-dash-dot-dash).
Vertical and horizontal axes have logarithmic scales.
}
\label{gamma_tronc_log_3_ngc2362}
\end{figure*}

A second  test  is performed 
on  low-mass stars in NGC 2362,
see  \cite{Irwin2008}.
Table
\ref{chi2valuesngc2362}  shows the statistical 
parameters which  
characterize the   astrophysical   fits.
Figure  \ref{gamma_tronc_log_ngc2362}  shows the fit
with  the  TG  distribution
of  NGC 2362
and Figure  \ref{gamma_tronc_log_3_ngc2362}
visually compares the three types of fits for NGC 2362.
\begin{table}[ht!]
\caption
{
Numerical values of
of the NGC 2362 cluster data (272 stars).
The  number of  linear   bins, $n$, is 20.
}
\label{chi2valuesngc2362}
\begin{center}
\begin{tabular}{|c|c|c|c|c|c|c|}
\hline
PDF       &   parameters  &  AIC  & $\chi_{red}^2$
& $Q$  &  D &   $P_{KS}$  \\
\hline
lognormal &  $\sigma$=0.5,$\mu_{LN} =-0.55 $  &  37.64&
1.86    & 0.013   &    0.07305  & 0.10486     \\
\hline
gamma & b=0.13 ,c =4.955 & 34.28 & 1.68   & 0.034
& 0.059  & 0.284   \\
\hline
truncated ~gamma & b=0.161 ,c =3.933
& 33.88 & 1.61   & 0.055  & 0.071  & 0.122   \\
~  & $x_l$=0.12, $x_u$=1.47
& ~     & ~      &  ~     &~       & ~        \\
\hline
four    &  Eqn.(58)
& 77.608 & 4.89    & $1.17\,10^{-8} $     & 0.16941
&  $2.6 \,10^{-7}$ \\
power~ laws  & in~Zaninetti~2013
&  ~            &   ~  & & & \\
\hline
\end{tabular}
\end{center}
\end{table}

\section{Conclusions}

The right or left TG PDF has been extensively
investigated in the field of mathematics , as  an example
\cite{Coffey2000} reports  most of the mathematical details.
The application of the TG PDF in astronomy
represents conversely a new promising field.
Here we have deduced the constant of normalization
 ,eqn.(\ref{constant}),
 the average value
 ,eqn.(\ref{meangammatruncated}),
the DF
, eqn.(\ref{dftruncatedgamma}),
and presented an algorithm  for the generation of the
random numbers , (eqn.\ref{randomgammatruncated}).
The application of the  TG PDF to the   IMF is positive
and both the reduced $\chi^2$ 
and  the K-S test 
give  better results in respect to the
standard PDFs used by the astronomers which are the lognormal and
the four power laws , see Tables  
\ref{chi2valuesngc6611} and  \ref{chi2valuesngc2362}.
A  comparison with the left truncated beta PDF , see Tables 1 and
2 in \cite{Zaninetti2013a} allows to say that the
left truncated beta PDF produces a better fit
to the IMF in respect to the truncated gamma PDF here analyzed.


\end{document}